# Josephson $\varphi_0$-junction in nanowire quantum dots

D. B. Szombati[1], S. Nadj-Perge[1], D. Car[2], S. R. Plissard[3], E. P. A. M. Bakkers[1,2], L. P. Kouwenhoven[1]

[1]QuTech and Kavli Institute of Nanoscience, Delft University of Technology, 2600 GA Delft, The Netherlands
[2]Department of Applied Physics, Eindhoven University of Technology, 5600 MB Eindhoven, The Netherlands
[3]CNRS-Laboratoire d'Analyse et d'Architecture des Systèmes (LAAS), Université de Toulouse, 7 avenue du colonel Roche, F-31400 Toulouse, France

**The Josephson effect describes supercurrent flowing through a junction connecting two superconducting leads by a thin barrier [1]. This current is driven by a superconducting phase difference $\varphi$ between the leads. In the presence of chiral and time reversal symmetry of the Cooper pair tunneling process [2] the current is strictly zero when $\varphi$ vanishes. Only if these underlying symmetries are broken the supercurrent for $\varphi$ = 0 may be finite [3–5]. This corresponds to a ground state of the junction being offset by a phase $\varphi_0$, different from 0 or π. Here, we report such Josephson $\varphi_0$-junction based on a nanowire quantum dot. We use a quantum interferometer device in order to investigate phase offsets and demonstrate that $\varphi_0$ can be controlled by electrostatic gating. Our results have possible far reaching implications for superconducting flux and phase defined quantum bits as well as for exploring topological superconductivity in quantum dot systems.**

The process of Cooper pair tunneling through a Josephson junction (JJ) is, in general, symmetric with respect to time inversion. This has a profound consequence for the JJ current-phase relation, $I(\varphi)$. In particular it imposes the condition $I(-\varphi) = -I(\varphi)$ which in turn results in $I(\varphi = 0)$ being strictly zero. The $I(\varphi = 0)=0$ condition is a consequence of the fact that for each process contributing to current flowing in one direction there is an opposite time reversed process, in which spin-up and spin-down electrons are reversed, that exactly cancels this current. However, time inversion is not the only symmetry which can protect the $I(\varphi = 0) = 0$ condition. For example, in JJs based on single domain ferromagnets, time inversion is broken but still the supercurrent is zero for $\varphi$ = 0 due to chiral symmetry, i.e. the symmetry between leftward and rightward tunneling. This symmetry assures that the tunneling coefficient describing the electron tunneling from the left lead to right lead is exactly the same as the one describing the tunneling vice versa, from the right lead to the left. The two tunneling processes (leftward and rightward) cancel each other which again results in $I(\varphi = 0)$ being strictly zero. This is even the case for so-called π-junctions[6] in which the current flow is reversed compared to usual JJs but still the underlying symmetries warrant zero current for $\varphi$ = 0. In order to create conditions for a non-zero supercurrent to flow at $\varphi$ = 0, both symmetries need to be broken [7]. Various ways were theoretically proposed to break the underlying symmetries and create $\varphi_0$-junctions, including ones based on non-centrosymmetric or multilayer ferromagnets [3,8], quantum point contacts [4], topological insulators [9,10], diffusive systems [11,12], nanowires [13,14] and quantum dots [5,15,16]. Alternatively, an effective built-in phase offset can be obtained by combining 0- and π- junctions in parallel [17,18]. However, no experimental demonstration of a $\varphi_0$-junction was reported until now.

In quantum dots (QDs), breaking of both symmetries can be achieved by the combination of an external magnetic field and spin-orbit interaction (SOI) [5,15,16]. A finite Zeeman splitting between spin-up and spin-down electrons breaks the time reversal symmetry. On the other hand, breaking of the chiral symmetry is more subtle. It requires an interplay between the SOI and the magnetic field and it can only occur when multiple orbitals are accessible for electron transport, see Fig. 1a. When an electron goes in and out from the QD via only one orbital (Fig. 1a, upper panel) the tunneling coefficient is exactly the same for the leftward and the rightward tunneling direction. In this case the chiral symmetry is preserved. If, however, the electron changes orbital within the quantum dot during the course of tunneling (Fig. 1a, lower panel), an extra phase factor is acquired in the process of orbital mixing. This phase factor, arising from the SOI enabled orbital mixing, depends now on the tunneling direction and it is different for the leftward and rightward tunneling process. As a consequence, the two processes cannot cancel each other and the chiral symmetry is broken. Although we discussed here the case of a single electron tunneling through the QD, the same argument holds for the breaking of the chiral symmetry in the tunneling of Cooper pairs (see supplementary information section 1 and Ref 5). Note that in this scenario both symmetries are explicitly broken by the combination of magnetic field and SOI [5].

The device geometry is shown in Fig. 1b and Fig. 1c. A single nanowire, made of Indium Antimonide (InSb), is contacted using Niobium Titanium Nitride (NbTiN) as a superconductor to make two JJs forming a quantum interference device (SQUID). We choose InSb nanowires due to their large spin-orbit coupling and $g$-factors both of which are important for breaking time inversion and chiral symmetry at relatively low magnetic fields [19,20]. Details of the wire growth and superconducting contact deposition were reported previously[21] (see also the methods section). Electrostatic gates below the wire are used to create a tunable quantum dot in the longer JJ and control the switching supercurrent of the shorter reference JJ [19] (Fig. 1c). All measurements are performed at a base temperature of $T$ = 20 mK in a 3-axis magnetic field where the flux through the SQUID is applied along the y direction (Fig. 1c). Standard quantum dot characterization, while the reference junction is pinched off, is used to determine the values of the charging ($E_C$) and orbital ($E_{orb}$) energies as well as $g$-factors. Depending on the confinement details and QD occupation number we find $E_C$ = 2 - 3 meV, $E_{orb}$ = 0.3 – 1.5 meV and $g$ = 40 - 50 (Fig. 1d). We identify small peaks around zero bias as an onset of superconductivity and estimate the induced superconducting gap in the QD to be $\Delta^*$ = 20 - 50 µeV (see supplementary information section 2).

First we measure the SQUID response in current bias for zero in-plane magnetic field (Fig. 2). Switching currents for the reference and quantum dot JJ, $I_{cref}$ and $I_{cQD}$, satisfy $I_{cref} \gg I_{cQD}$, ensuring that the phase drop is mainly across the QD. The measured voltage as a function of flux and bias current $I_{bias}$ shows oscillations with a period of $B_Y$ = 1.2 mT (Fig. 2a) corresponding to an effective area of 1.8 µm², which is consistent with the SQUID geometry and the penetration depth of NbTiN ($\lambda \approx$ 170 nm). Both junctions are in the phase diffusive regime such that no hysteresis is observed (Fig. 2a, right panel). This allows probing of the phase response by applying a finite $I_{bias}$ = 100-500 pA close to $I_{cref}$ and monitoring the voltage drop across the SQUID, $V$, as a function of gate voltage $V_2$ and flux $\Phi$, see the lower panels in Fig. 2a and Fig. 2b as well as section 2 of the supplementary information.

In this QD regime, the phase of the SQUID pattern depends crucially on the dot occupation number (Fig. 2b). For example, for $V_2 \approx$ -247 mV, the measured voltage oscillates as a function of $\Phi$ with a particular phase (purple colored line in Fig. 2b lower panel). When $V_2$ is increased to around -240 mV the oscillations disappear and the overall voltage drops as the charge degeneracy point is reached. By increasing $V_2$ further, the oscillations recover with an extra π phase corresponding to the sign reversal of the supercurrent in a QD [22] (light blue line in Fig. 2b lower panel). The change of phase by π is repeated for several consecutive charge states.

The change in phase measured for zero in-plane field occurs due to the change in the electron parity of the ground state. In a simple physical picture, for odd QD occupancy, the order of electrons forming a Cooper pair is reversed in the process of co-tunneling through a single quantum dot orbital. This results in the sign reversal of the supercurrent and the observed π shift, as previously reported in Ref. 22. Note, however, that even if the phase of the ground state is changed, $I(\varphi =0)$ remains zero which is anticipated since time reversal symmetry is preserved.

Finite magnetic fields can substantially modify this simple picture in two ways. First, the QD levels split by Zeeman energy which results in different co-tunneling rates for spin-up and spin-down electrons and therefore breaks time reversal symmetry. Second, the spin split levels belonging to different orbitals move closer in energy which enables more than one orbital to contribute to the co-tunneling process. This in turn, combined with strong SOI induced orbital mixing and asymmetry in the barriers, results in the breaking of chiral symmetry (see supplementary information section 1) [16]. Under these conditions one can expect shifts in the phase by an arbitrary $\varphi_0$.

For finite in-plane magnetic fields we find regimes in which the shifts of the SQUID pattern are different from 0 or π. Instead, the shifts take non-universal values depending on the specific QD configuration and magnetic field direction and strength (Fig. 3). Fig. 3a and Fig. 3b show an example taken close to the QD charge degeneracy point. The shift in SQUID response between the two Coulomb blockade regions is approximately 0.7π. This value is considerably different from the value π observed for the same QD regime when the in-plane field is zero (compare the data in Fig. 3b and Fig. 3d with the data in Fig. 2b). Note also that while effects related to finite temperature have impact on the critical current values and in general on the values and visibility of the SQUID response they do not contribute to any phase offset (see supplementary information section 3).

The measured gate tunable phase shift directly implies a finite $\varphi_0$, different from 0 or π, for at least one of the Coulomb blockade regions. Importantly, this shift cannot be explained by simple higher harmonic terms in the JJ current-phase relation which can occur in various semiconductor based junctions [23–26]. Even if such terms were present, as long as $I(-\varphi) = -I(\varphi)$, the SQUID response would have to be symmetric around the points corresponding to integer values of the threaded flux. Since this is clearly not the case in the data shown in Fig. 3, the $I(-\varphi) = -I(\varphi)$ condition is violated. Note that both junctions in the SQUID are nanowire based and therefore phase shifts can occur in the reference junction as well. For this reason shifts in the SQUID pattern should be interpreted as relative offsets in $\varphi_0$ of the QD based junction.

Typically, the phase of the SQUID oscillation is constant within the Coulomb blockade region and changes only at the charge degeneracy points. Depending on the exact gate settings the phase change appears either as a discrete jump or a continuous transition. In the investigated regimes, we measured jumps when the QD is strongly confined (as in Fig. 3a and Fig. 3b). For a more open QD we observe a continuous change in the phase of the SQUID response as we tune the gate $G_2$ across the charge degeneracy point (Fig. 3c and Fig. 3d). This behavior is not fully understood but we note that transport for a strongly confined QD is dominated by the resonant tunneling process at the Coulomb peak and therefore can be very different compared to the transport deep in the blockaded regime. This effect is not pronounced for open QD in which higher order tunneling processes are relevant. In the regimes where the SQUID oscillations can be detected along the whole charge transition we observe a continuously changing phase. Importantly in all regimes fields of $B_{in-plane}$ ≈ 50 - 150 mT are required to see a noticeable shift in the SQUID response (see supplementary information section 4). These fields are still around two to four times smaller compared to the critical fields of $B_{\text{in-plane}}$ = 200-300 mT at which the SQUID response vanishes.

Finally, we examine the magnetic anisotropy dependence of the SQUID pattern, in order to further study the microscopic origin of the $\varphi_0$-junction. The data showing phase shifts between neighboring charge states for various in-plane magnetic field angles is presented in Fig. 4. Consistently, for many different QD regimes, we observe that the maximum shift of the SQUID pattern is most pronounced when an in-plane field is applied orthogonal to the nanowire. Previous quantum dot experiments have identified this field orientation with the preferential spin-orbit direction $B_{\text{SO}}$ for quantum dots [20]. These measurements are consistent with SOI enabled orbital mixing which predicts maximal phase $\varphi_0$ for $B_{\text{in-plane}}$ || $B_{\text{SO}}$ [5,15,16]. Note that other known mechanisms which could in principle lead to additional phase shifts, such as flux penetrating the JJ area, are not consistent with the observed data (see supplementary information section 5 for a more detailed discussion).

In summary, we demonstrated a gate tunable Josephson $\varphi_0$-junction. Results presented here imply that the breaking of the underlying symmetries can be achieved in superconductor-quantum dot structures while maintaining coherent transport of Cooper pairs. In this context, our experiment is directly related to the efforts of studying triplet superconductivity as well as in achieving topological superconducting phase in quantum dots coupled to an s-wave superconductor [16,27–31]. Aside from that, a gate tunable phase offset may open novel possibilities for the realization of electrically controlled flux and phase based quantum bits[32] as well as superconducting "phase" batteries and rectifiers [4,33]. Finally, we note that that other one-dimensional materials, such as carbon nanotubes where spin-orbit is strong due the curvature of the tube, may be explored in the context of $\varphi_0$-junctions[34].

Acknowledgments: We gratefully acknowledge Raymond Schouten, Sergey Frolov, Dmitry Pikulin, Attila Geresdi, Kun Zuo, Vincent Mourik, Anton Akhmerov, Michael Wimmer, Y. Nazarov, and C. Beenakker for useful discussions and their help. This work has been supported by funding from the Netherlands Foundation for Fundamental Research on Matter (NWO/FOM), Microsoft Corporation Station Q and the ERC synergy grant.

**Methods:**

The Indium Antimonide (InSb) wires used in the experiments were grown using MOVPE process. Before superconducting contacts deposition wires were etched in Ar$^+$ plasma for 120 seconds to remove native surface oxides. NbTiN was sputtered in similar conditions as in Ref. 14. Finite offsets in the SQUID response corresponding to $\varphi_0$-junction were observed in three separate cooldowns of the device.

**Author contributions:**

D.B.S. fabricated the sample and D.B.S. and S. N-P performed the measurements. D.C., S.R.P. and E.P.A.M.B. grew the InSb nanowires. All authors discussed the data and contributed to the manuscript.

**Figure captions:**

**Figure 1. Schematics of the experiment. a,** Schematics showing tunneling of an electron through the QD with two orbitals labelled 1 and 2 which are mixed by the SOI. The blue (red) line describes tunneling of an electron from the left (right) to the right (left) lead. When there is no change in orbital the two processes cancel each other (upper panel). In contrast when the orbital is changed during the tunneling (lower panel), due to interplay between the SOI and a magnetic field B, leftward and rightward tunneling processes do not cancel. In this case an extra phase $\chi$ is obtained in the process, which depends on the strength of the SOI and on $B_{in\text{-}plane}$. Note that the phase for forward and backward tunneling are opposite. **b,** Device schematic showing a dc-SQUID measured in a four terminal geometry. Voltages $V_1$, $V_2$, $V_3$, and $V_{ref}$ are applied on underlying gates to control the conductance of the JJs. **c,** Scanning electron microscopy (SEM) image of the actual device. Gates $G_1$, $G_2$ and $G_3$ are used to define a quantum dot in the long JJ while $G_{ref}$ tunes the current through the reference JJ. Orientation of the in-plane magnetic fields $B_X$ and $B_Z$ are marked. $B_Y$ is used for tuning the flux $\Phi$ through the SQUID. **d,** Current as a function of $V_2$ and $B_X$ showing QD evolution of the Coulomb peak spacing in the field which gives g-factor $g_x \approx 51$. From similar data taken for $B_Z$ we obtain $g_z \approx 44$ and estimate the level repulsion between different spin configurations due to SOI of $\Delta_{SO} \approx 170$ µeV. The extracted $\Delta_{SO}$ corresponds to a spin-orbit length $l_{so} \approx 350$ nm and a spin-orbit energy $E_{SO} \approx 20$ µeV comparable to previous QD experiments [19]. Measurements are performed in the voltage bias regime, $V_{bias} = 500$ µV. The dashed rectangle indicates the range of $B_X$ for which the $\varphi_0$-junction is observed.

**Figure 2. Nanowire SQUID characterization for $B_{in\text{-}plane} = 0$. a**, Voltage across the SQUID, $V$, as a function of bias current $I_{bias}$ and flux $\Phi$ through the SQUID. The right panel shows $V$ vs $I_{bias}$ measured at $\Phi = 4\,\Phi_0$ (cut along the orange dashed line). The switching current $I_S$ separating low and high resistance regions is indicated. The lower panel shows voltage vs $\Phi$ for $I_{bias} = 450$ pA (cut across the green dashed line). **b,** $V$ as a function of $V_2$ and $\Phi$ for $I_{bias} = 190$ pA. The phase of the SQUID oscillations is alternating between 0 and π depending on the electron parity of the ground state of the QD. The right panel shows Coulomb peaks in the voltage bias regime. The bottom panel shows $V$ vs. flux cuts at $I_{bias} = 195$ pA for $V_2 = -247$ mV (purple) and $V_2 = -233$ mV (light blue).

**Fig 3. Observation of a continuous phase change in the Josephson $\varphi_0$-junction for finite $B_{in\text{-}plane}$. a,c,** $V$ as a function of $V_2$ ($V_3$ in panel **c**) and Flux at fixed current bias ($I_{bias} = 470$ pA, $B_{in\text{-}plane} = 120$ mT and $\theta = -135°$ for panel **a**; $I_{bias} = 240$ pA, $B_{in\text{-}plane} = 75$ mT and $\theta = -35°$ for panel **c**). Here $\theta$ is the angle between the direction of the in-plane magnetic field and the nanowire axis. In contrast to the data taken at zero in-plane magnetic field, the phase shift of the voltage oscillations in flux is tunable with gate voltage $V_2$ ($V_3$ in panel **c**). **b,d,** $V$ vs flux for values of $V_2$ ($V_3$ in panel **d**) marked by dashed lines on panels **a** and **c** showing phase shifts. In panel **b** the black curve is taken at $V_2 = -285$ mV and $I_{bias} = 460$ pA and the orange at $V_2 = -240$ mV and $I_{bias} = 470$ pA. The relative offset from the two curves is $0.35 \pm 0.1\,\Phi_0$. In panel **d** the curves are cuts from panel **c** taken at $V_3$ values of 213 mV; 218 mV; 225 mV; 229 mV. The corresponding

offsets in phase compared to the top curve are (0.1 ± 0.05) $\Phi_0$, (0.3 ± 0.05) $\Phi_0$ and (0.4 ± 0.05) $\Phi_0$. Note that in the QD regime shown in panel **c** and **d** we used gate $G_3$ for tuning.

**Fig. 4 Anisotropy of the SQUID phase shift for various angles of $B_{\text{in-plane}}$. a-e**, Voltage vs. flux for different orientations of $B_{\text{in-plane}}$ = 120 mT. Red and blue curves in each panel are taken at two neighboring charge occupations as in Fig 3a and the corresponding relative phase shift between is marked above each panel. The maximum shift from $\pi$ was obtained when the field is perpendicular to the wire as expected from the SOI enabled orbital mixing (see also supplementary information section 4). **f**, Phase offset as a function of angle $\theta$ between the nanowire and $B_{\text{in-plane}}$.

**Figures:**

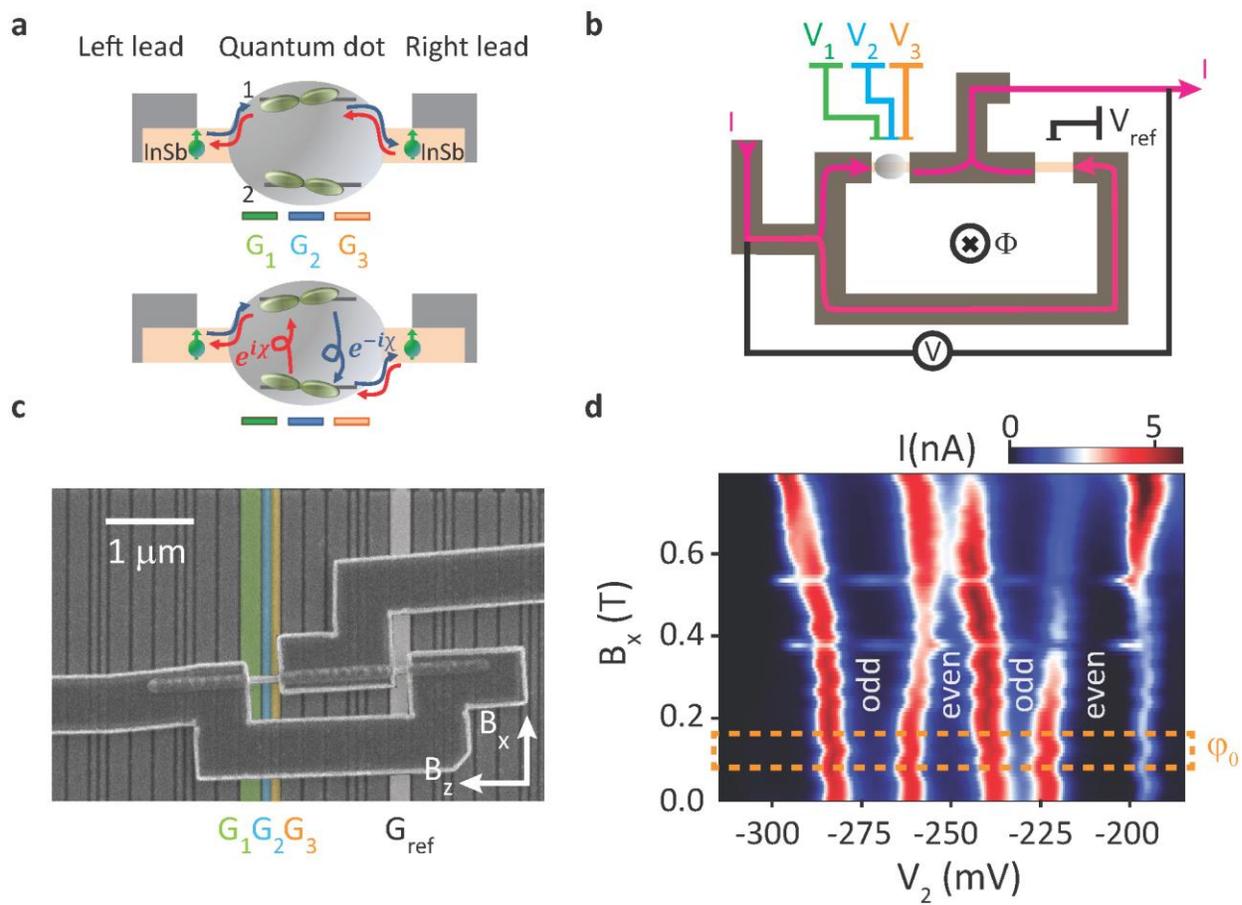

**Fig. 1**

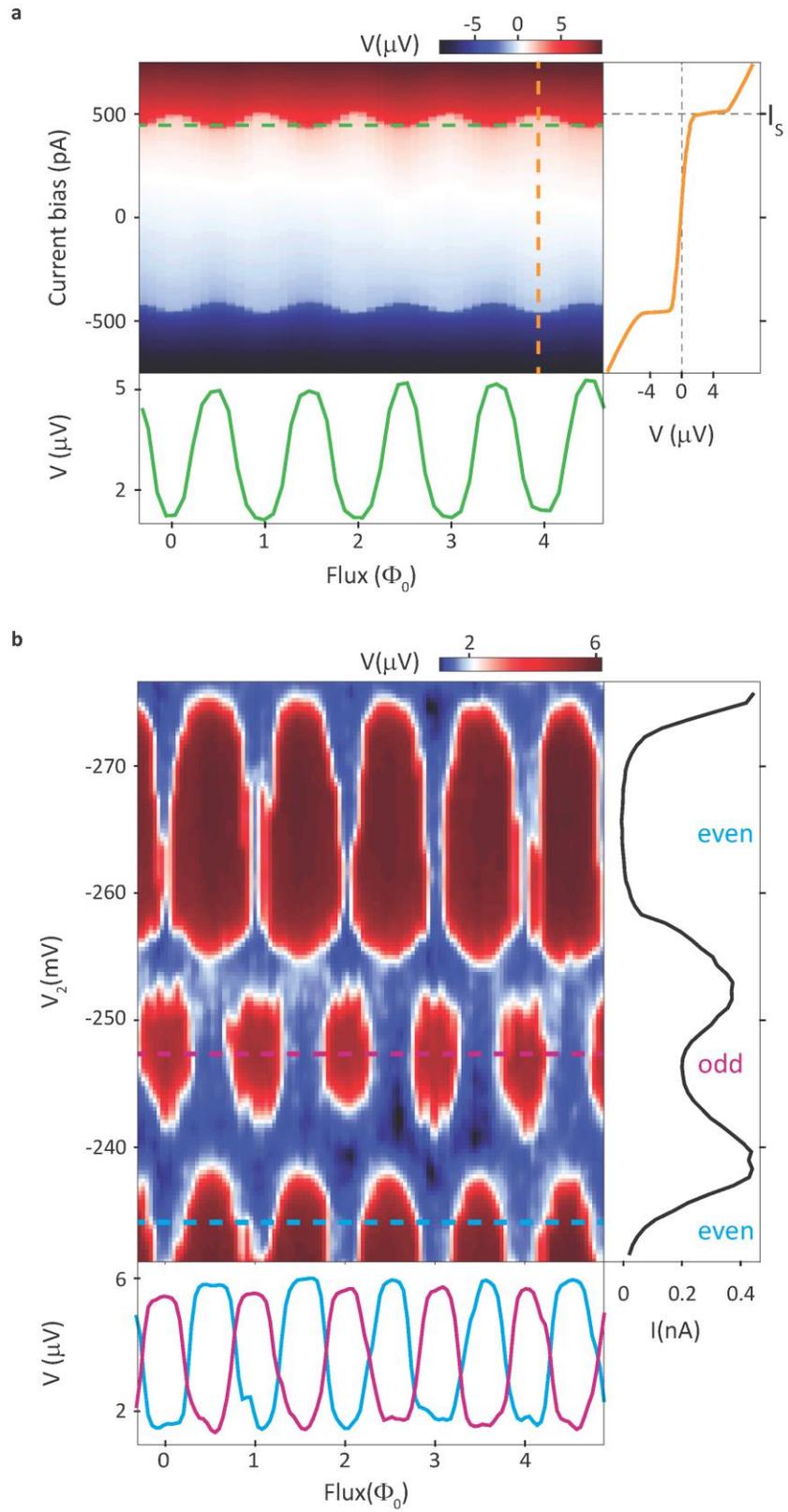

**Fig. 2**

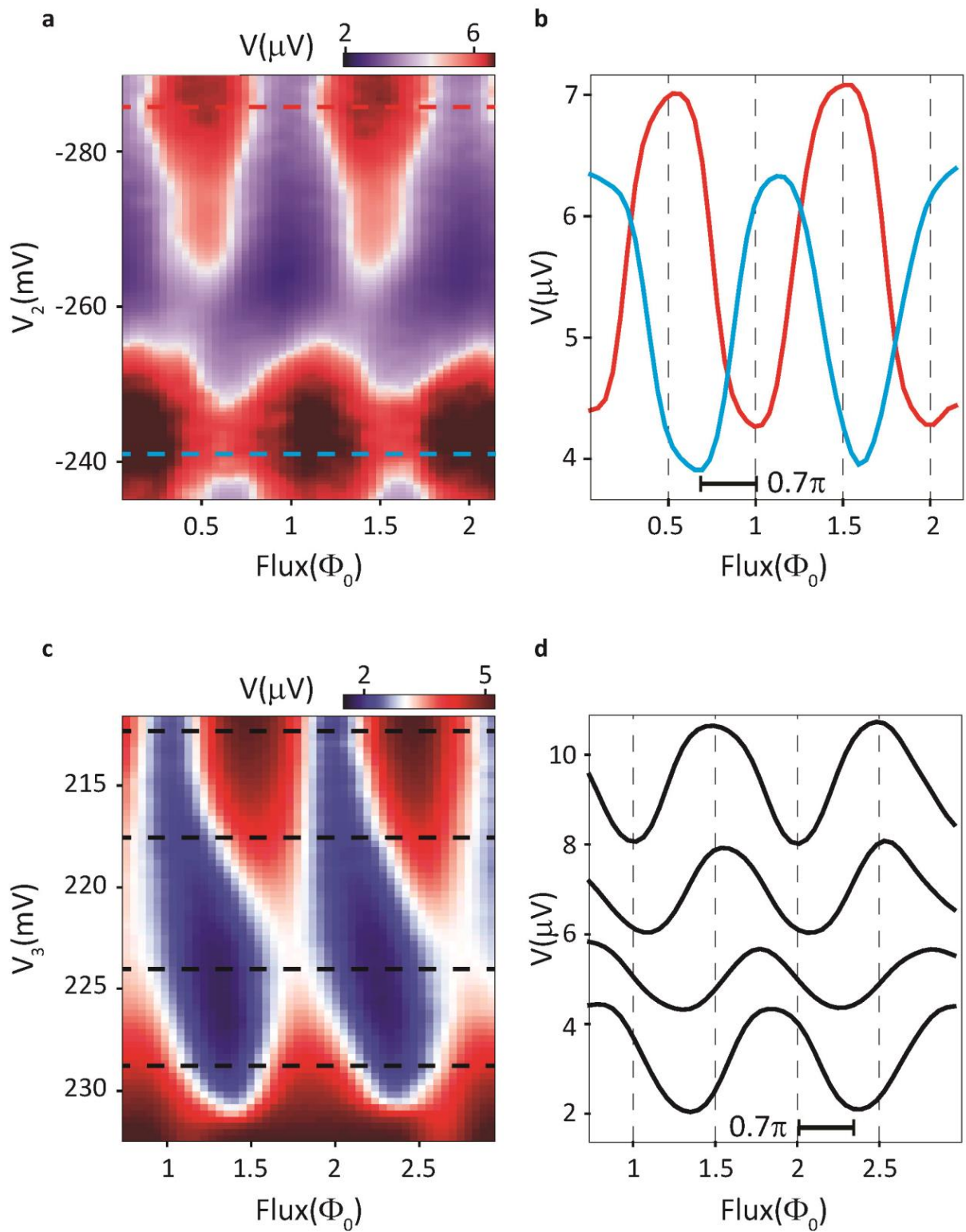

**Fig. 3**

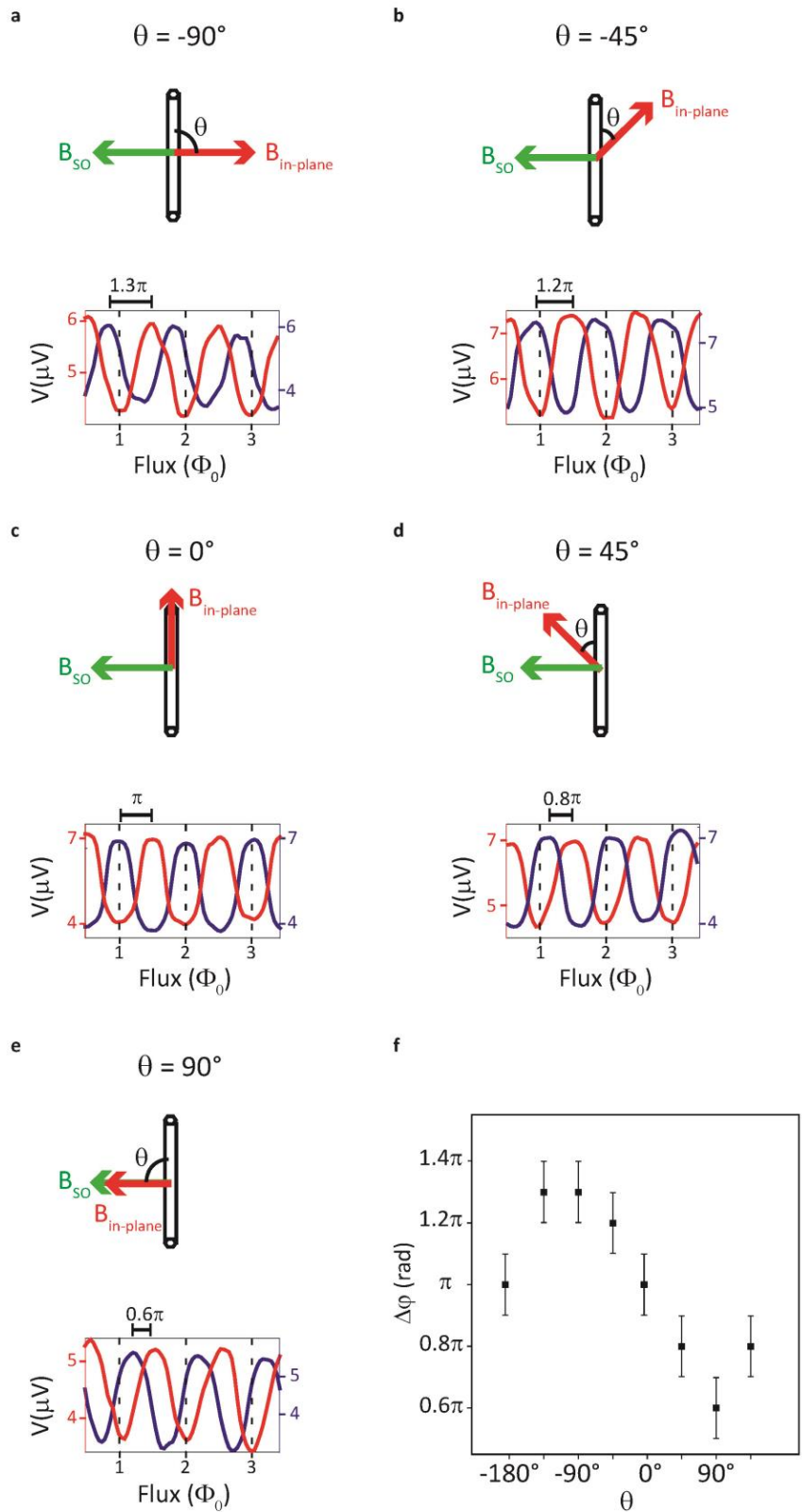

**Fig. 4**

**1. Breaking of the chiral symmetry in quantum dots**

**2. Characterization of the quantum dot junction and the nanowire based SQUID**

**3. Shifts of the SQUID phase pattern**

**4. Additional data**

**5. Establishing the origin of the shift in the SQUID pattern**

**6. Anomalous current and direction dependent critical current in $\varphi_0$-junctions**

**7. Estimation of the anomalous current**

# 1. Breaking of the chiral symmetry in quantum dots

In one-dimensional systems in which the electron momentum is well defined, the interplay between the spin-orbit interaction (SOI) and the Zeeman splitting can create a difference between the dispersion of electrons moving forward and backward. This in turn can lead to the breaking of the chiral symmetry and, in the case of superconducting transport, to Josephson $\varphi_0$-junctions [1–3]. In quantum dots (QDs) there is no well-defined momentum since the QD states are localized. Nevertheless, the combination of the SOI and the external magnetic field still creates similar conditions for breaking of the chiral symmetry as shown in Refs. [4–6] and discussed below.

Let us consider a process describing a Cooper pair tunnelling from the left to the right lead (rightward tunnelling) at zero phase difference. Without SOI electrons forming the Cooper pair tunnel through the QD via a single orbital level, for example the first electron tunnels via level 1 and the second via level 2. The corresponding tunnelling coefficient (matrix element) for this process is given by $(t_{L1}t_{R1})(t_{L2}t_{R2})$. Here the $t_{L1}$ and $t_{L2}$ ($t_{R1}$ and $t_{R2}$) are the hybridization amplitudes between QD levels 1 and 2 with the left (right) lead. The terms in brackets correspond to tunnelling coefficients for individual electrons. Assuming that the hybridization amplitudes are real, the matrix element describing tunnelling from the right to the left (leftward tunnelling) is exactly the same. Since the leftward tunnelling contributes to the current flow in the opposite direction, the net resulting current vanishes. Therefore, the tunnelling via single orbitals can not add to $I(\varphi = 0)$. The lowest order process which contributes to $I(\varphi = 0)$ is the one in which one electron tunnels through the dot directly via a single orbital, while the other electron changes the orbital during the tunnelling process. Finite SOI enables such orbital change.

In the simplest case when two quantum dot levels contribute to Cooper pair transport and the magnetic field is orientated along the effective spin-orbit axis the Hamiltonian of the dot can be written as

$$H_{QD} = (\mu\tau_0 + E_{orb}\tau_Z)\sigma_0 + B\tau_0\sigma_Z + \alpha\tau_Y\sigma_Z \quad (1)$$

Here $\mu$ is the chemical potential, $E_{orb}$ is the orbital energy, $\alpha$ parametrizes the strength of the SOI and $B$ the Zeeman splitting, $\tau_{X,Y,Z}$ ($\sigma_{X,Y,Z}$) are Pauli matrices acting in orbital (spin) space ($\tau_0$ ($\sigma_0$) are identity matrices). Usually the terms describing the Zeeman splitting and the SOI are smaller in comparison to the first term in the Hamiltonian. In the presence of SOI the eigenstates of the QD are mixtures of the two orbital states. This mixture between QD eigenstates and the left (right) lead can be expressed in terms of the single level tunnelling coefficients as $t_{L(R)1'} = t_{L(R)1} \cos \varepsilon + i \sin \varepsilon \, t_{L(R)2}$ and $t_{L(R)2'} = t_{L(R)2} \cos \varepsilon - i \sin \varepsilon \, t_{L(R)1}$ for spin-up electrons (with $\sin \varepsilon = \alpha/E_{orb}$). For the spin down electrons + and - signs should be inverted.

Importantly, due to orbital mixing, the coefficients describing tunnelling events become complex numbers implying that electrons crossing the junction gain a finite phase. This phase is opposite for the electrons tunnelling in the other direction. Therefore the rightward and leftward tunnelling coefficients are not exactly the same (the imaginary part is different) and the two tunnelling processes do not cancel each other. If Cooper pairs also acquire a finite phase during the tunnelling process, $I(\varphi = 0)$ becomes finite. However, if the magnetic field is zero, since spin-up and spin-down electrons obtain the opposite phases in the tunnelling process, Cooper pairs do not gain phase even when SOI is present. For finite magnetic fields the tunnelling probabilities for the tunnelling of spin-up and spin-down electrons via different orbitals are no longer exactly the same. Only in this case can Cooper pairs obtain a finite phase.

Finally we stress that the complex tunnel coupling between superconductors (occurring due to the combination of SOI and finite magnetic field) always leads to finite $I(\varphi=0)$. Interestingly, this follows even from Feynman's simplified description of the Josephson effect [7]. If we assume that the wavefunctions describing the two superconductors are $\psi_L = |\psi_L|e^{i\varphi_L}$ and $\psi_R = |\psi_R|e^{i\varphi_R}$, the time dependent Hamiltonian describing the superconductors on the two side of the junction can be written as

$$i\hbar\frac{\partial}{\partial t}\psi_L = \mu_L\psi_L + T\psi_R$$

$$i\hbar\frac{\partial}{\partial t}\psi_R = T^*\psi_L + \mu_R\psi_R.$$

Here $\mu_L$ and $\mu_R$ are the chemical potentials in the two superconductors and $T$ is the tunnel coupling. Solving this set of equations for current directly gives

$I \sim |\psi_L||\psi_R|$ (Re($T$) sin($\varphi_L - \varphi_R$) + Im($T$)cos($\varphi_L - \varphi_R$)).

When $T$ is real current is proportional to sin($\varphi$), with $\varphi = \varphi_L - \varphi_R$. However if the imaginary part is non-zero, the term ~ cos ($\varphi$) also contributes to the current and gives rise to the finite $I(\varphi=0)$.

**2. Characterization of the quantum dot junction and the nanowire based SQUID**

In order to characterize the QD Josephson junction, we performed voltage and current bias measurements while the reference junction was pinched off. Depending on the exact gate configuration, the measured QD resistance varies between 40-600 kΩ and the switching currents are in the range 40-300 pA. In all measurements the sub-gap resistance is finite since the Josephson energy of the QD junction $E_J = \Phi_0 I_C / 2\pi \approx$ 0.5-3 μeV is comparable to $k_BT \approx$ 5 μeV. The induced gap in the QD junction is of the order of 20-50 μeV (see Fig. S2 (**b**), (**d**)).

When the reference junction is open we observe standard SQUID oscillations. In this regime it is even easier to resolve small supercurrents of the QD junction by simply estimating the amplitude of the flux dependent voltage oscillations. Note that the data presented in the main text is taken with the SQUID tuned to the overdamped regime. However, at low magnetic fields, the SQUID is usually underdamped (Fig. S3). Due to hysteresis effects, in this case, phase offsets are difficult to track in the voltage vs flux measurements when the current bias is fixed. For this reason, before each measurement we made sure that SQUID is in the overdamped regime by tuning the switching current of the reference junction via $G_{ref}$.

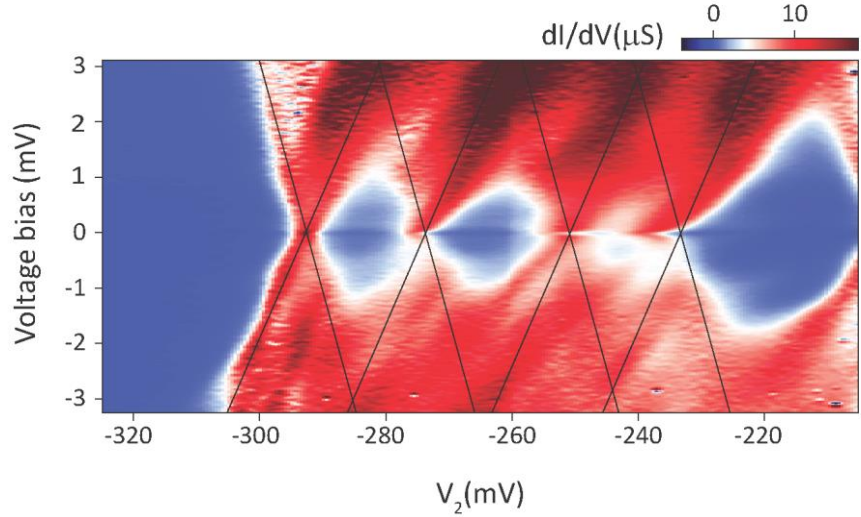

Fig. S1: Coulomb blockade diamonds for the same gate configuration as in Fig. 1 and Fig. 2 ($V_1$ = 350 mV; $V_3$ = 110 mV).

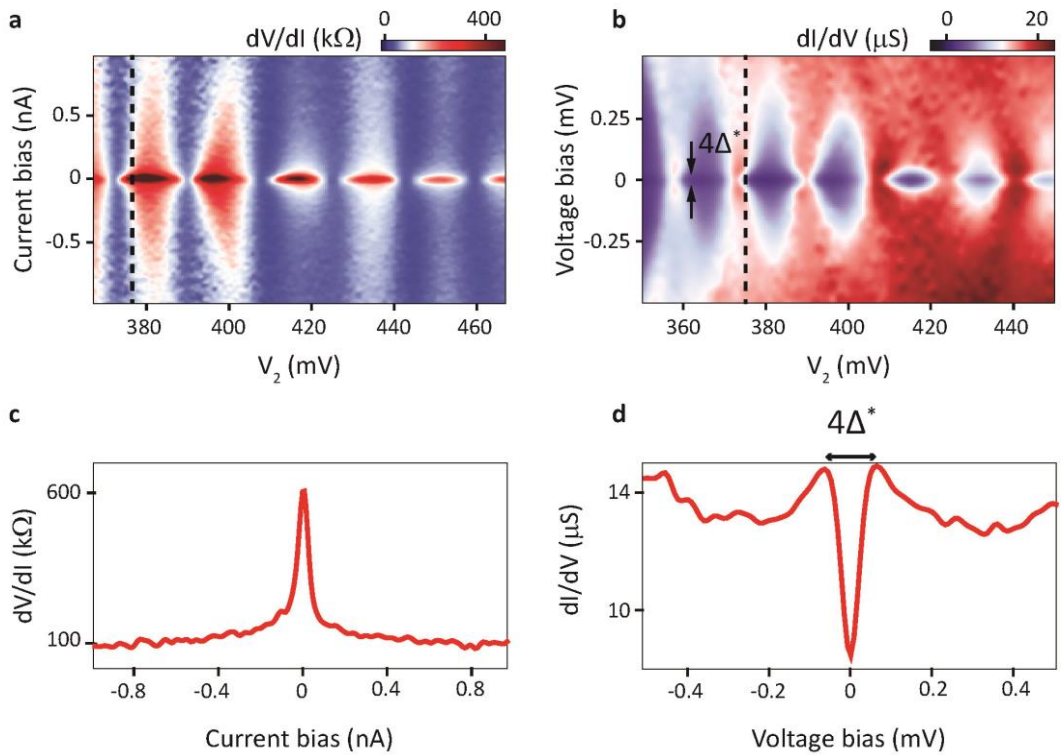

Fig. S2: Coulomb blockade in the current biased (**a**) and the corresponding voltage biased (**b**) regime ($V_1$ = 350 mV; $V_3$ = 110 mV) with $V_2$ being more positive. In this regime QD has ~30 electrons more compared to Fig. S1. **c,d**, Linecuts for current (voltage) bias for the fixed gate voltage showing resistance (conductance) of the QD junction. The sudden increase in resistance corresponds to the suppression of the density of the states inside of the superconducting gap $\Delta^* \approx$ 25 µeV in this regime. Note that the coupling between the QD and the leads is larger compared to $\Delta^*$.

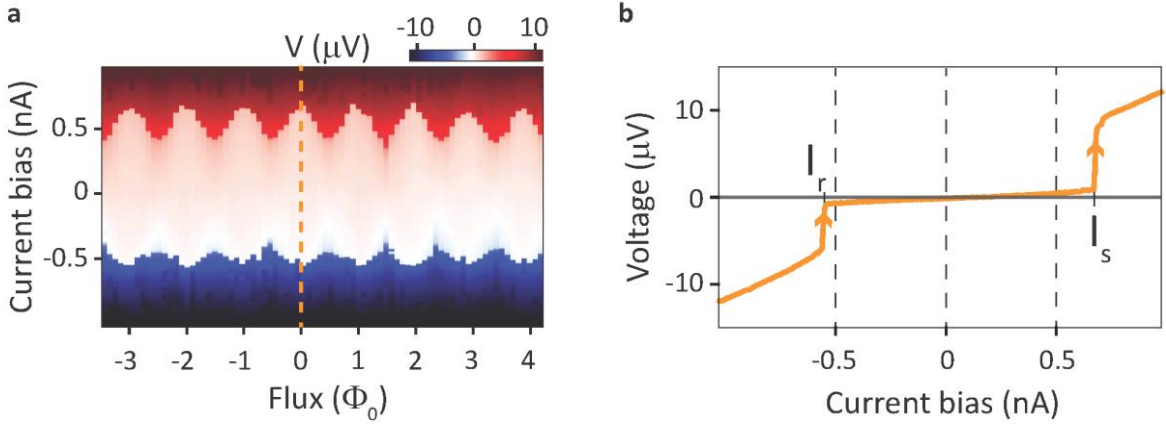

Fig. S3: SQUID in the underdamped regime at zero in-plane field ($V_1$ = 100 mV; $V_2$ = 25 mV; $V_3$ = 335 mV; $V_{ref}$ = 420 mV). **a**, Voltage as a function of flux and the current bias showing hysteresis effects in the switching and retrapping current. **b**, Line cut along the dashed line in (**a**) showing a difference of around 200 pA between switching and retrapping currents.

### 3. Shifts of the SQUID phase shift pattern

In order to understand the origin of the shifts in our SQUID patter, we have performed simulations of the critical current of a dc-SQUID consisting of two Josephson junctions with current-phase relationships (CPR) $I_{S_{1,2}} = \text{CPR}_{1,2}(\varphi_{1,2})$, where $I_{S_1}$ ($I_{S_2}$) is the switching current of junction 1 (junction 2) and $\varphi_1$ ($\varphi_2$) is the superconducting phase difference across junction 1 (junction 2).

Assuming negligible SQUID inductance, the phase difference across the junctions are related to each other by the equation $\varphi_2 = \varphi_1 - 2\pi \frac{\Phi_{ext}}{\Phi_0}$, where $\Phi_{ext}$ is the external flux applied through the SQUID. The critical current of the SQUID is the calculated using the equation

$$I_{C,SQUID} = \max_{\varphi} \left| \text{CPR}_1(\varphi) + \text{CPR}_2\left(\varphi - 2\pi \frac{\Phi_{ext}}{\Phi_0} + 2\pi n\right) \right|.$$

In Fig. S4 (**a**) we plot the critical current of the SQUID assuming $I_{S_{1,2}} = I_{C_{1,2}} \sin\varphi_{1,2}$ for several values of the ratio $\frac{I_{C_1}}{I_{C_2}}$. We observe that although the shape of the $I_{C,SQUID}$ vs flux curve varies, the points in flux of the maxima and minima are fixed. In Fig. S4 (**b**) we plot the same curves but now $\text{CPR}_1$ is a periodic sawtooth, i.e. $I_{S_1} = I_{C_1}\left(\frac{1}{\pi}\varphi - 1\right)$. Still the minima and the maxima of the critical current remain fixed, independent of the ratio $\frac{I_{C_1}}{I_{C_2}}$.

In Fig. S4 (**c**) we assume $I_{S_1} = 0.1 I_{C_2} \sin(\varphi_1 - \varphi_0)$ and vary the value of $\varphi_0$. In contrast to the previous two cases, the maxima and minima of the critical current now shift by $\Phi_{ext} = \frac{\varphi_0}{2\Pi}$. Considering that the maximum (minimum) of $I_{C,SQUID}$ corresponds to a minimum (maximum) in the measured voltage V over the SQUID, such $I_{C,SQUID}$ behaviour is qualitatively similar to the measured voltage pattern shown in Fig 3. (**b**), (**c**) and Fig. 4 of the main text as well as Fig S6 (**b**) of this supplementary. In our simulations the only way we could induce additional phase shifts in the SQUID pattern is to offset one of the junctions by $\varphi_0$. Therefore these are consistent with our interpretation that the origin of the shift in our measured SQUID patterns is indeed a consequence of the shift by $\varphi_0$ in CPR of the nanowire Josephson junction.

To be certain that the magnitude of the critical currents does not influence the phase of the SQUID oscillations, we repeated scans with the same gate voltages over the quantum dot $V_1$, $V_2$, $V_3$, and varied $I_{C_{ref}}$ by changing the gate voltage of the reference junction $V_{ref}$ over a wide range and found no change of the induced shift, even when the SQUID is underdamped regime.

This is understandable since the phase difference between the junctions in a SQUID geometry is fixed by the external flux. So while the relative ratio between the critical currents (which is not exactly known due to phase diffusion) have impact on the critical current values and in general on the visibility of the SQUID response they do not change the phase offset. This fact is illustrated in the experiment of *Spathis et al.* [8] , where the authors measured the nanowire SQUID response in the wide range of temperatures and found no shifts in the pattern.

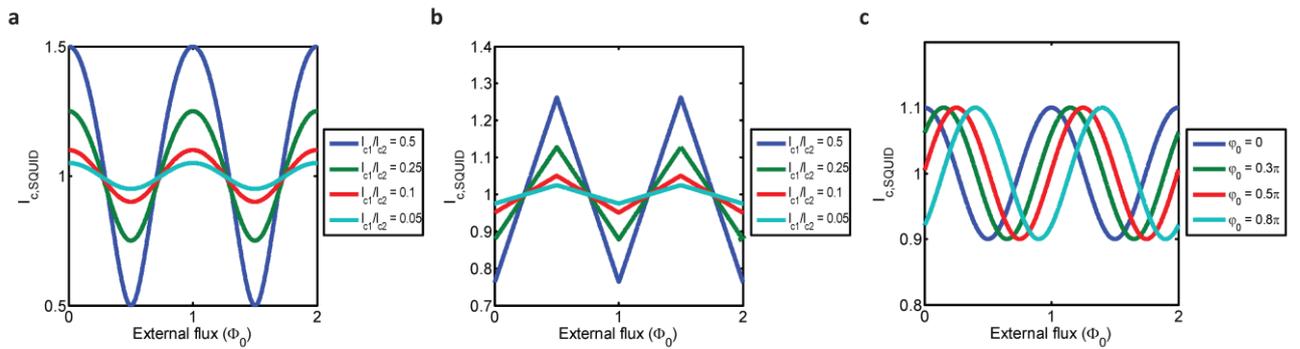

Fig. S4: Critical current simulations of a dc-SQUID. **a**, Simulation including two sinusoidal junctions with varying critical current ratios. **b**, Simulation including a junction 1 with a sawtooth-like CPR and junction 2 with a sinusoidal CPR, with varying critical current ratios. **c**, Simulation including two sinusoidal junctions where the CPR of junction 1 is shifted by a phase $\varphi_0$. The ratio $\frac{I_{C_1}}{I_{C_2}} = 0.1$.

## 4. Additional data

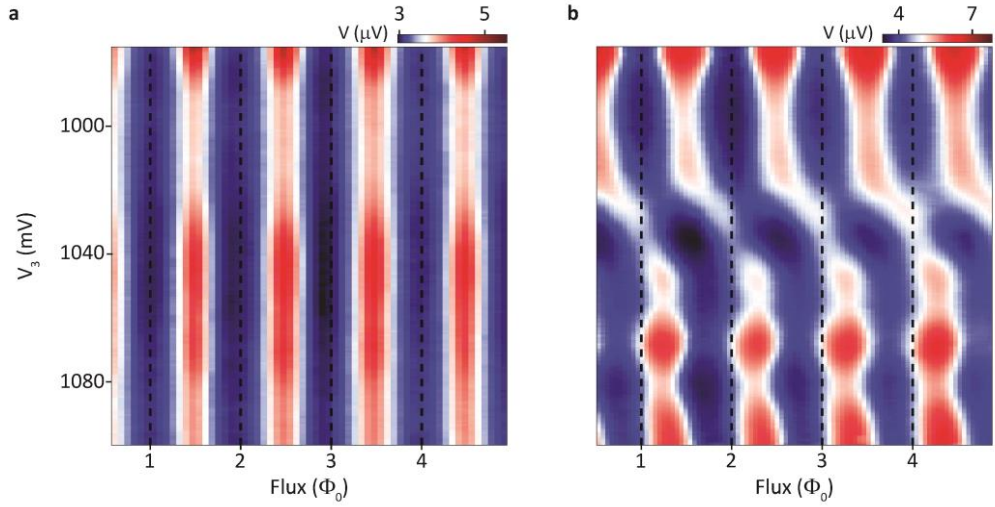

Fig. S5: Measured voltage as a function of flux and $V_3$ ($V_1$ = 100 mV; $V_2$ = 50 mV; $I_{bias}$ = 220 pA; $V_{ref}$ = 450 mV;) **a**, $B_{in-plane}$ = 0. **b**, $B_{in-plane}$ = 150 mT, $\theta$=75°. In this regime no 0-π transition is observed suggesting that multiple quantum dot orbitals contribute to the transport. The phase shifts are mainly constant inside the regions of gate space in which the quantum dot occupation number is fixed.

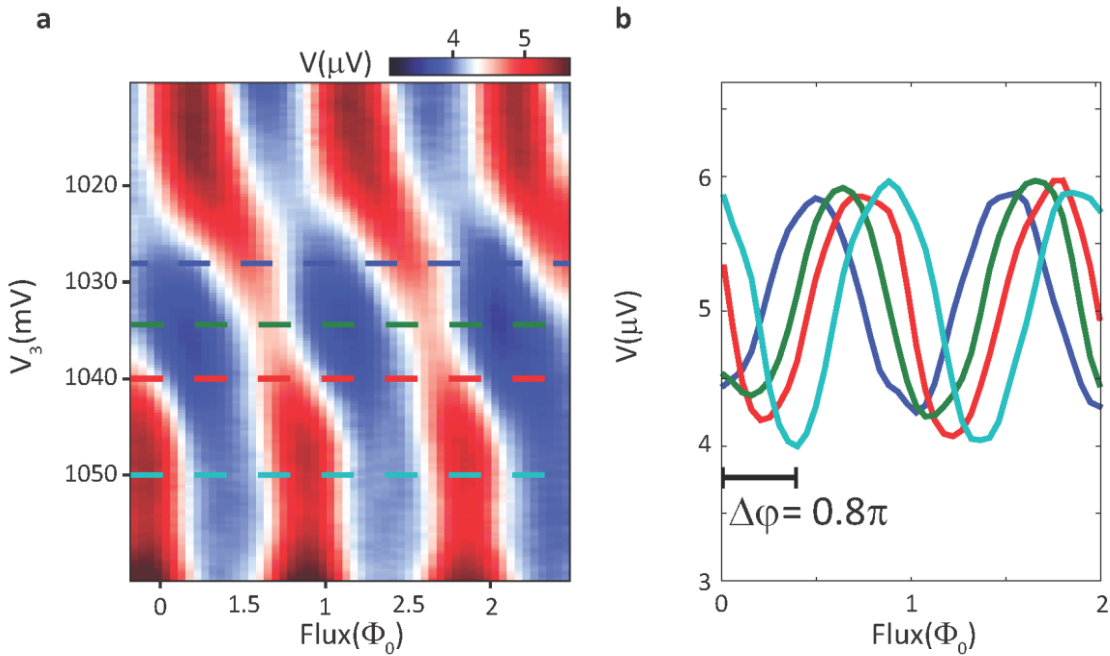

Fig. S6: Continuously gate tunable $\varphi_0$-shift. **a:** zoom in on Fig. S5 (**b**). The dashed lines represent the values of $V_3$ at which the curves in (**b**) are taken. **b**, Measured voltage vs flux taken at consecutive $V_3$ values marked in (**a**). The total phase shift between the blue curve and cyan curve is $0.8\pi$.

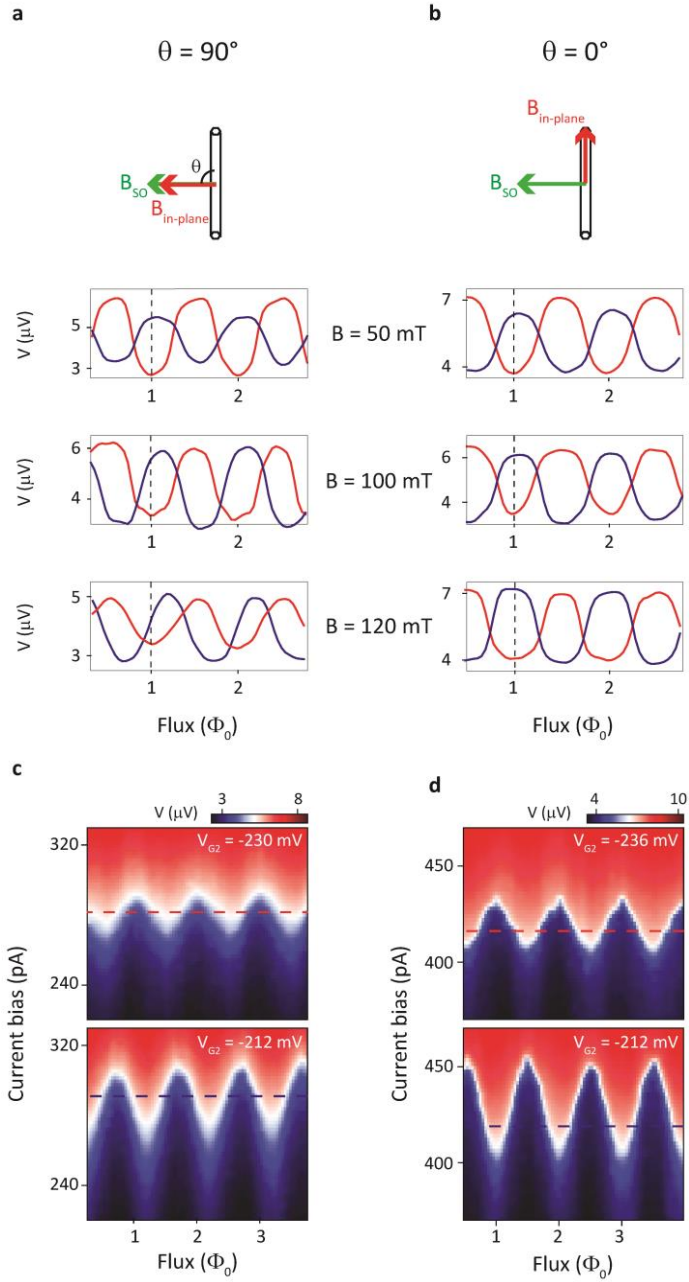

Fig. S7: Evolution of the shift in the SQUID pattern with the magnetic field for two different magnetic field orientations: **a**, orthogonal to the nanowire. **b,** Along the nanowire. The blue and red traces correspond to the two consecutive quantum dot occupation states. **c,d**, Voltage as a function of flux and current bias at $B_{\text{in-plane}}$ = 120 mT for the same field orientation as in (**a**) and (**b**). The sharp transition from the low voltage state (blue) to the high voltage state (red) indicates the value of the switching current as a function of flux. The phase offset is independent of the current bias. The red and blue lines correspond to the current bias at which the data in the lowest panel of (**a**) and (**b**) is taken.

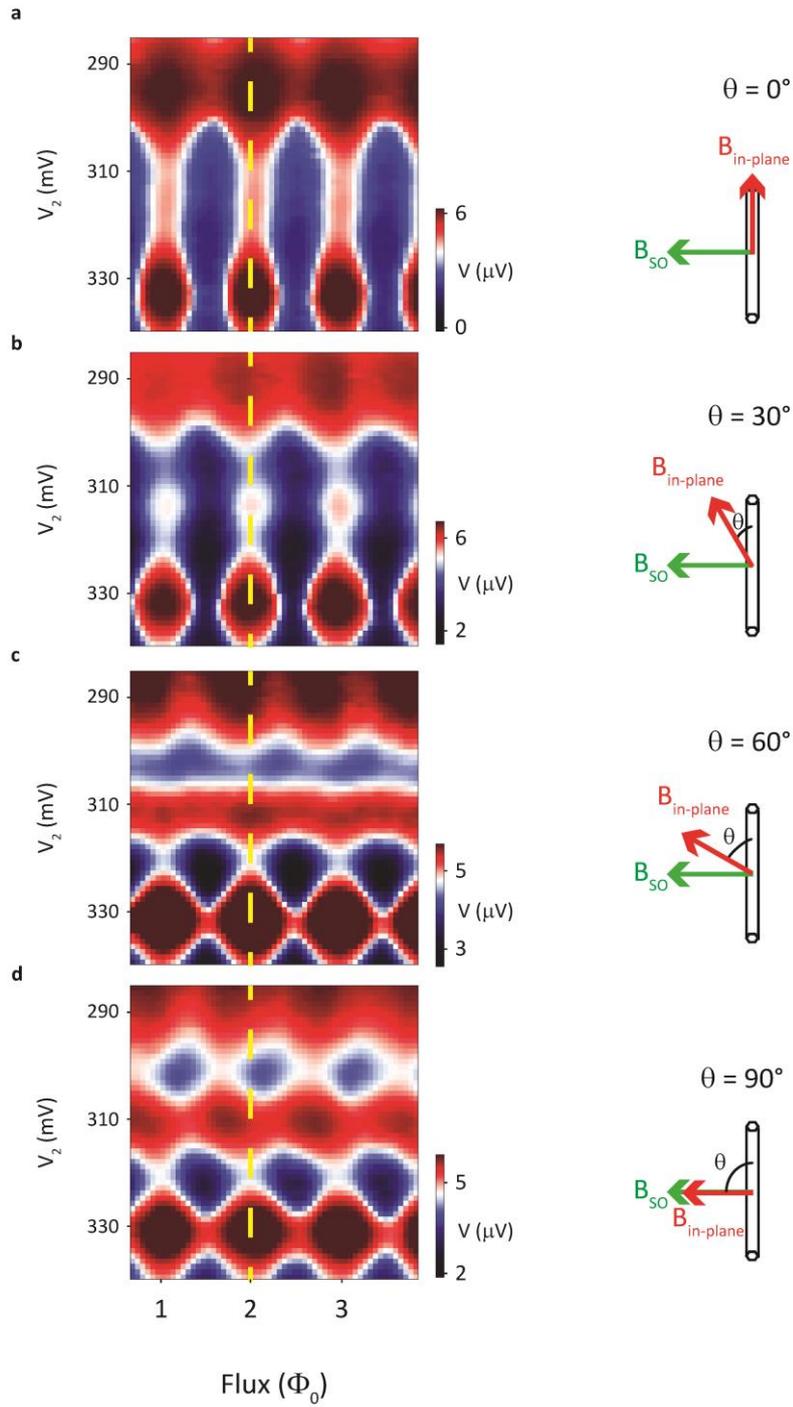

Fig. S8: **a-d**, Anisotropy of the SQUID phase shift in the open QD regime. Voltage as a function of $V_2$ and flux for different orientations of $B_{in-plane}$ and **30-50** more electrons compared to the regime in Fig. 4. Here the 0-π transition was not observed strongly suggesting that multiple orbitals are contributing to the transport. In this very different regime compared to the data discussed in the main text the $\varphi_0$ shifts are still the largest when the external in-plane field ($B_{in-plane}$= 100mT) is oriented orthogonal to the nanowire.

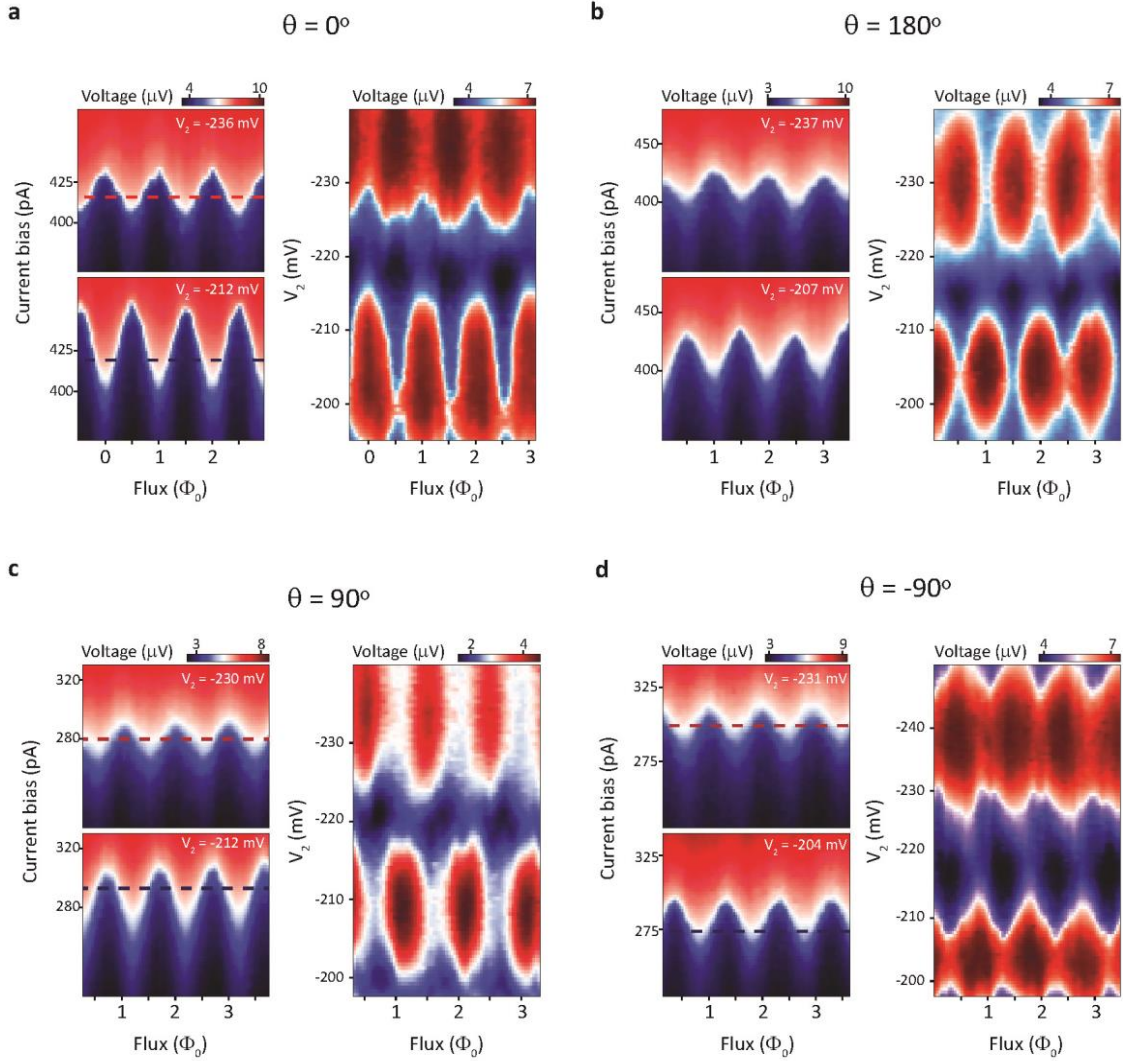

Fig. S9: Additional anisotropy data for $B_{\text{in-plane}}$ = 120 mT. **a-d**, The left panel shows voltage vs current bias and flux for the gate settings corresponding to two consecutive Coulomb blockade regions. The right panel shows voltage vs V$_2$ and flux. The angle $\theta$ between the nanowire and **B**$_{\text{in-plane}}$ is indicated. Blue and red dashed lines indicate cuts shown in Fig. 4. The corresponding values of the $I_{\text{bias}}$ are: (**a**) top panel $I_{\text{bias}}$ = 415 pA , bottom panel $I_{\text{bias}}$ = 420 pA; (**c**) $I_{\text{bias}}$ = 280 pA, $I_{\text{bias}}$ = 290 pA; (**d**) $I_{\text{bias}}$ = 295 pA, $I_{\text{bias}}$ = 275 pA.

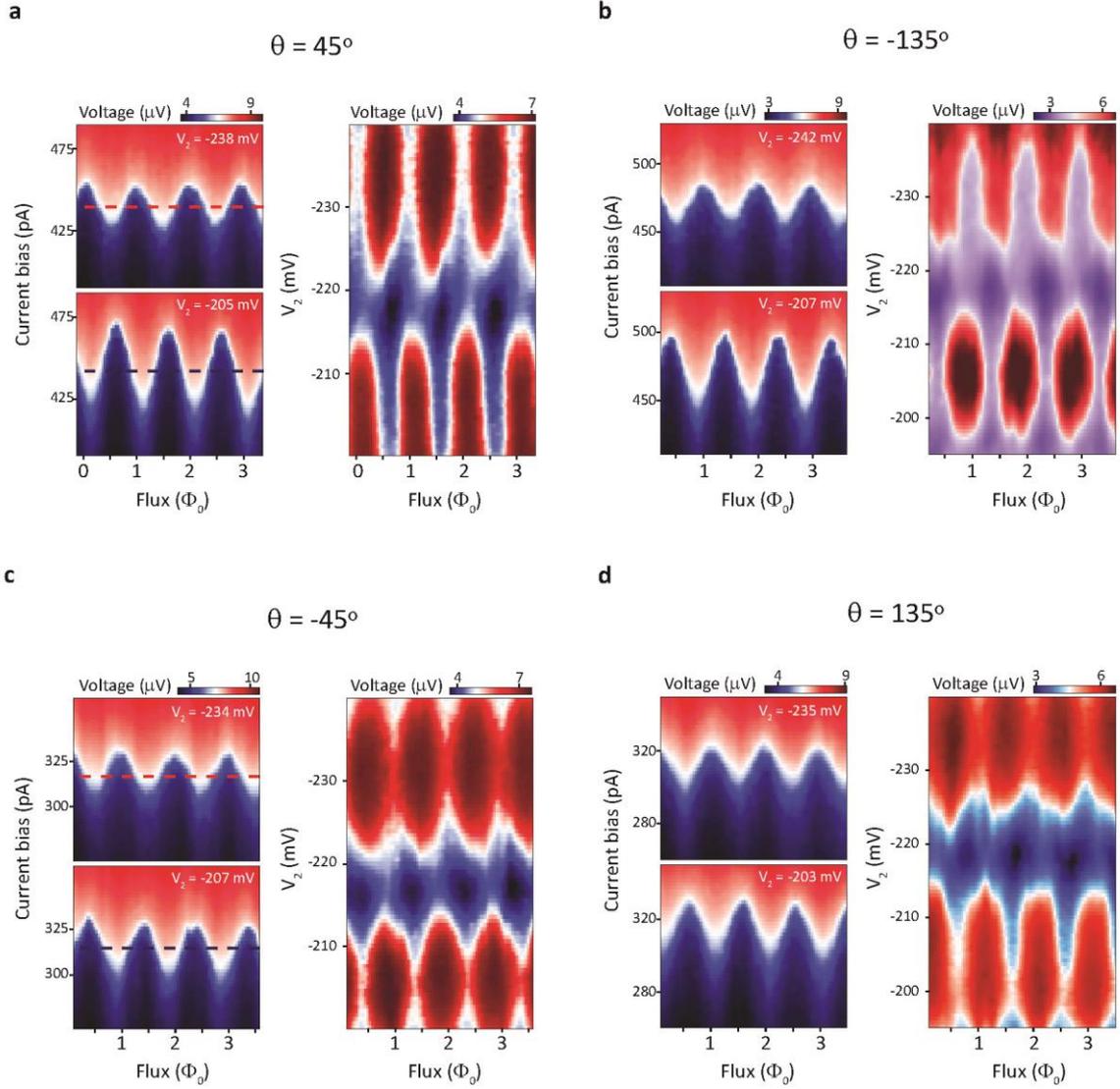

Fig. S10: Additional anisotropy data for $B_{\text{in-plane}}$ = 120 mT. **a-d**, The Left panel shows voltage vs current bias and flux for the gate settings corresponding to two consecutive Coulomb blockade regions. The right panel shows voltage vs $V_2$ and flux. The angle $\theta$ between the nanowire and $B_{\text{in-plane}}$ is indicated. Blue and red dashed lines indicate cuts shown in Fig. 4. The corresponding values of the $I_{\text{bias}}$ are: (**a**) top panel $I_{\text{bias}}$ = 440 pA, bottom panel $I_{\text{bias}}$ = 440 pA; (**c**) $I_{\text{bias}}$ = 320 pA, $I_{\text{bias}}$ = 315 pA.

## 5. Establishing the origin of the shift in the SQUID pattern

Our main experimental observations can be summarized as follows: (1) the observed shift in the SQUID pattern occurs for a finite in-plane magnetic field which exact value depends on the QD configuration; (2) the shift in pattern occurs mainly for gate values at which the QD electron occupation number changes; (3) the shift is the largest when the field is orthogonal to the nanowire and almost non-existing when the field is oriented along the nanowire.

These observations are qualitatively in agreement with SOI induced orbital mixing as the origin of the $\varphi_0$–junction. Based on (1) and (2) it is evident that QD orbital levels play a crucial role in the superconducting transport which is also in agreement with previous experiments on quantum dots [9–11]. Also, the observed anisotropy is consistent with reported SOI direction in QDs [12]. In the following we discuss other effects which may also contribute to the observed shifts in the SQUID pattern.

**a) Gate induced changes in the effective SQUID area.** Gating off a part of the wire changes the effective SQUID area which may result in additional shifts of the interference pattern. This effect is rather small in our devices. The maximal change in the area, and therefore the phase offset, would be at most few percent estimated by comparing the gated nanowire area 100nm x 100nm with the total area of the SQUID. Even if assumed that the magnetic field is enhanced in the vicinity of the nanowire junction, due to complicated field profile caused by the nearby superconductor, the change in area has to be extremely large to account for the observed shift. Also, for substantial changes in the area, the flux periodicity of the SQUID response has to change substantially. These changes were not observed in the experiment which shows periodicity of 1.2mT being independent of the gate parameters. We also note that we didn't observe any discontinuous jumps in the interference pattern while sweeping the magnetic field which rules out phase shifts due to accidental events of flux trapping in the junction.

**b) Phase offsets due to flux in the quantum dot.** The observed shifts in the SQUID pattern were obtained in in-plane field values of 50-100 mT. Assuming the quantum dot area to be 60nm x 60nm (corresponding to $E_{orb}$ = 1.5 meV), the total flux through the corresponding area would be of the order of 0.1-0.2 $\Phi_0$. Based on this estimate, even if the flux through the QD would fully add to the $\varphi_0$ offset, the resulting shift would be too small to explain the experimental data. Note that we verified that there is no significant modification of the field profile in the vicinity of the quantum dot by measuring the values of the g-factors.

**c) Additional orbital effects.** As discussed in Ref. [5,6], when the tunnelling coefficients (matrix elements) describing the hybridization between the QD levels and the left (right) lead are complex numbers and contribute to additional phase factors, at finite magnetic field an anomalous Josephson current may occur. Orbital effects can also contribute to this complex phase instead of SOI. While this is in principle a possible scenario it is not consistent with the experimental data as one would not expect any magnetic field anisotropy in this case. Another possible scenario is that orbital effects alone can result in an anomalous current. As pointed out in Refs. [13,14] orbital effects alone may have a significant influence on superconducting transport through the nanowire without any QD. Although these effects may indeed contribute, they are to the large degree linear in magnetic field strength in contrast to the experimental data. For this reason, we can rule out these effects as the main contribution of the observed shifts.

## 6. Anomalous current and direction dependent critical current in $\varphi_0$-junctions

The subject of $\varphi_0$-junctions has been theoretically extensively studied in the past. They have been predicted to arise in many different systems besides quantum dots [1,5,16], such as conventional superconductors with spin-orbit coupling [17–19], with triplet correlations [20–22], superconductors in contact with topological materials [23,24] and also hybrid systems with nonconventional superconductors [25–27].

The current-phase relation (CPR) for conventional Josephson junctions states that the switching current varies with the sine of the phase difference across the junction: $I_S(\varphi) = I_0 \sin\varphi$, where the junction's critical current $I_C = I_0$. This CPR can be generalized by adding a cosine term:

$$I_S(\varphi) = I_0 \sin\varphi + I_{anomalous} \cos\varphi \equiv I_C \sin(\varphi + \varphi_0),$$

where the critical current is now expressed as $I_C = \sqrt{I_0^2 + I_{anomalous}^2}$. For conventional 0-junctions ($\varphi_0 = 0$) and π-junctions ($\varphi_0 = \pi$, $I_S(\varphi) = -|I_0|\sin\varphi \equiv I_0 \sin(\varphi + \pi)$), the anomalous term vanishes and there is no current flowing when the phase difference $\varphi = 0$. This can be seen in Fig. S11 (**a**) where we plot the switching current as a function of the phase difference φ for a 0-junction and a π-junction.

It follows that a $\varphi_0$ term different from 0 or π directly implies the existence of a finite anomalous current. In Fig.S11 (**b**) we show the anomalous current for a junction with $\varphi_0 = 0.15\pi$. This is a shifted sine curve, hence $I_{C+} \equiv \max_\varphi I_S = I_{C-} \equiv \left|\min_\varphi I_S\right|$, meaning that the critical current is independent of the bias direction. In order to be able to measure a different critical current when the bias is reversed, i.e. to satisfy the condition $I_{C+} \neq I_{C-}$, the CPR needs to contain higher order terms, e.g. as in the experiment by *Sickinger et al.* [15] In Fig. S11 (**c**) we plot the switching current for a junction with $I_S(\varphi) = \sin(\varphi + 0.35\pi) - 0.5\sin 2\varphi$ and indeed obtain two different critical currents as shown on the plot.

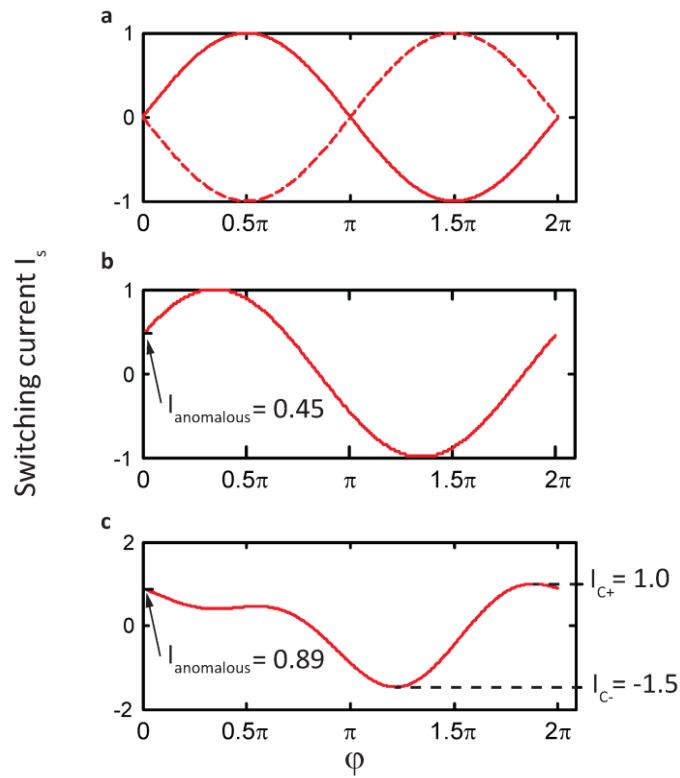

Fig. S11: Switching current for junctions with various CPR normalized to $I_0$. **a**, $I_S(\varphi) = \sin\varphi$ (continuous line) and $I_S(\varphi) = \sin(\varphi + \pi)$ (dashed line). **b**, $I_S(\varphi) = \sin(\varphi + 0.15\pi)$, $I_{anomalous} = I_S(0)$ is shown. **c**, $I_S(\varphi) = \sin(\varphi + 0.35\pi) - 0.5\sin 2\varphi$, anomalous current and direction dependent critical current is shown.

## 7. Estimation of the anomalous current

Using the data from the regime presented in Fig. S7 (**a**) and the relation $I_{anomalous} = I_C \sin \varphi_0$, we estimate the minimum anomalous current through our quantum dot.

The procedure we use the estimate the anomalous current goes as following. We assume that the critical current $I_C$ is constant along charge state transitions for a fixed **B**$_{in\text{-}plane}$. Within a charge transition where we measure a relative change in phase of $\varphi_0$, we choose the larger value of the possible magnitude of the anomalous current, i.e. $\max(|I_C \sin \varphi|, |I_C \sin(\varphi + \varphi_0)|)$. However, since the phase difference $\varphi$ across the quantum dot is unknown, we minimize this function over all possible values of $\varphi$. Thus we obtain our minimum estimate of the magnitude of the anomalous current $|I_{\min\_anomalous}|$ as

$$|I_{\min\_anomalous}| = \min_{\varphi \in [0,\pi]} \{\max(|I_C \sin \varphi|, |I_C \sin(\varphi + \varphi_0)|)\}.$$

$|I_{\min\_anomalous}|$ vanishes for $\varphi_0 = 0, \pi$ as expected for zero and $\pi$-junctions, and is positive otherwise. Fig. S12 shows the values extracted for particular $\varphi_0$-shifts measured. Note that shift is non-linear function of the field. While for a small magnitude values of **B**$_{in\text{-}plane}$ the anomalous current is negligible, above certain critical value of the field the shift abruptly increases.

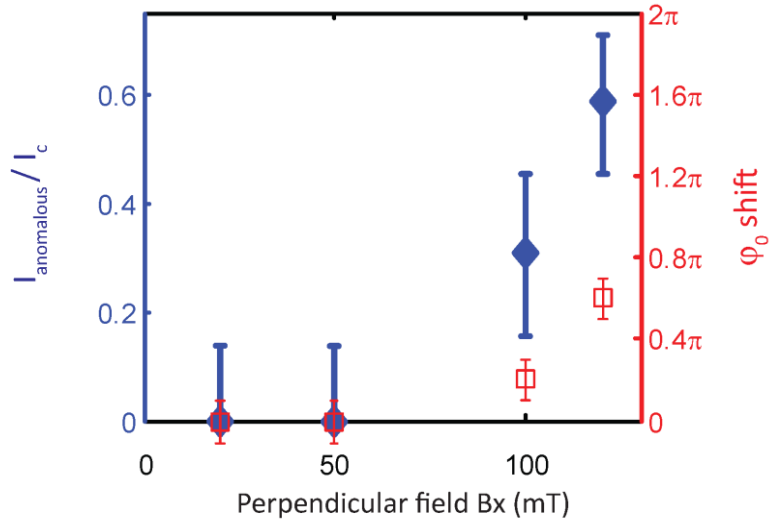

Fig. S12: Anomalous current estimates and measured phase shifts as a function of field magnitude for the regime shown in S7 (perpendicular in-plane field).